\documentclass[print,epsfig,onecolumn,floats,showpacs]{revtex4}
\usepackage{graphicx}
\usepackage{dcolumn}
\usepackage{bm}
\usepackage{amssymb}
\usepackage{amsmath}
\usepackage{epsfig}
\usepackage[titletoc]{appendix}
\begin{document}
\author{Xi-wang Luo, Yong-jian Han, Guang-can Guo, Xingxiang Zhou}
\email{xizhou@ustc.edu.cn}
\author{Zheng-wei Zhou}
\email{zwzhou@ustc.edu.cn}
\title{\textbf{Simulation of non-Abelian Anyons using ribbon operators connected to a common base site}}
\begin{abstract}
A convenient and effective way in the quantum double model to study anyons in a topological space
with a tensor product structure is to create and braid anyons
using ribbon operators connected to a common base site [A.~Kitaev Ann.\ Phys. (N.Y.) \textbf{303}, 2 (2003)].
We show how this scheme can be simulated in a physical system by constructing long ribbon operators
connected to a base site that is placed faraway. We describe how to move and braid anyons
using these ribbon operators, and how to perform measurement on them. We also give the smallest scale of a system
that is sufficient for proof-of-principle demonstration of our scheme.

\end{abstract}
\maketitle

\section{Introduction}

Anyons are exotic quasi-particles in two dimensional systems that obey fractional
statistics \cite{F.W}. By the associative properties of their underlying algebra, they are divided
into Abelian and non-Abelian anyons. Under particle exchange, the wavefunction of Abelian
anyons acquires a phase that can be different than multiples of $\pi$. For non-Abelian anyons,
the wavefunction is subject to a none diagonal unitary gate when two particles
are exchanged \cite{F.W,Wen}.
Anyons are important for understanding the physics of two-dimensional strongly
correlated systems and they are speculated to exist
in fractional quantum Hall fluids \cite{GS,CZG1,CZG2}.

Aside from their fundamental implications in many-body physics, in recent years anyon systems
have been suggested as a promising candidate for realizing intrinsically robust
quantum computation because of their topological properties ~\cite{A.K1}. For instance, Kitaev proposed
two exactly solvable spin lattice models with anyonic excitations, the quantum double model and the
honeycomb model \cite{A.K1,A.K2}. These schemes are called topological quantum computation
(TQC), and they take advantage of the
topological invariance of anyon qubits to protect against local noise.

In spite of the conceptual significance of anyons and their appeal for quantum computation
applications, it is very difficult to study anyons experimentally because the conditions for
their existence and observation are extremely challenging to realize. To date, there has only
been some experimental evidence in support of the existence of Abelian anyons in fractional
quantum Hall fluids \cite{GS,CZG1,CZG2}. For the purpose of universal TQC, Abelian anyons are
insufficient and non-Abelian anyons must be available. There have been several
theoretical proposals for direct observation of non-Abelian anyons in quantum Hall systems
\cite{DFN,BKS} and the honeycomb model \cite{DDL,MBZ,Zhang}. Unfortunately, these proposals require very
complex setup or very low temperatures that are beyond the reach of current experimental
capabilities.

Considering the difficulty in realizing physical systems with genuine anyonic particles,
quantum simulation of anyons is very valuable since it provides a viable alternative for
studying the kinematics of topological states. In this practice, we will not try to
construct the complex many-body Hamiltonian giving rise to genuine anyons which often
involves interactions between more than two physical particles. Rather, the goal is to
find practical methods to create topological states in a realizable system and study
their properties. Though these states are not truly topologically protected, they
can be created, manipulated and observed in a properly designed simulation system, and
therefore are very valuable for the study of anyon physics and topological quantum
computing.

Recognizing the importance of quantum simulation of anyon physics,
in recent years researchers have proposed schemes for simulating
both the Abelian \cite{HRD,Pan} and non-Abelian
\cite{ABVC1,ABVC2,ABVC3} quantum double model based on photonic
and trapped atomic systems. Though these schemes are very valuable
for the research of anyon physics, they also have serious
limitations. In particular, the issue of using topologically
protected manipulations only has not been sufficiently addressed.
For instance, local fluxes and local charges are used in
\cite{ABVC2} to encode quantum information and these local degrees
of freedom will be disturbed by local noise such as local gauge
transformations. In \cite{Wooton}, though information in the
quantum memory is encoded with anyon states, the computing
operations are non-topological. Therefore, it is necessary to
study new schemes that can demonstrate genuine fractional
statistics and simulate topologically protected quantum computing.

In order to simulate anyon states that can truly demonstrate
non-Abelian statistics and universal TQC, we must carefully
examine the topologically protected space in the system which is
the computational Hilbert space for TQC. In general, this
topologically protected space does not have a tensor product
structure and unitary transformations of states in it under
braiding cannot be described clearly. Consequently, it is unclear
how non-Abelian statistics is manifested in this space and how it
can be used for computation \cite{A.K1}. To overcome this
difficulty, Kitaev suggested that one can use an arbitrary but
fixed \textit{base site} as a reference \cite{A.K1} to create
anyons. By constructing appropriate ribbon operators that share
the chosen base site as one of their ends (see Fig. \ref{fig
ribbon} (a)), one can create anyonic excitations on the other end
of the ribbons. Since all these anyonic excitations have the same
reference base site, an overall topologically protected space can
be constructed as a tensor product of the topologically protected
space associated with each ribbon operator \cite{A.K1}. The
corresponding Hilbert space does have a tensor product structure
and the transformations of anyon states in it under braiding and
fusion can be easily derived. Kitaev's scheme offers a feasible
and effective method for performing topologically protected
quantum computation. Unfortunately, a scheme to directly simulate
this important approach of Kitaev's has not been available, since
it is nontrivial to construct appropriate ribbon operators
required for its realization in a physical system.

In this work, we show how to dynamically simulate the $S_3$
non-Abelian quantum double model by proposing a scheme to realize
ribbon operators connected to a common base site $x_{0}$. We
describe how to create anyonic excitations using these ribbon
operators, and how to move the other end of the ribbon to achieve
the braiding of anyonic excitations on it. We also demonstrate how
to detect the braiding and fusion, and propose a method to realize
anyonic interferometry using controlled ribbon operators to detect
the topological states of the anyonic excitations. With these
capabilities, we can then study the fractional statistics of
non-Abelian anyons in the quantum double model, and also simulate
universal TQC. Based on this, we investigate the minimum scale of
a system that is required for proof-of-principle demonstration of
our scheme, and propose an implementation using superconducting
circuits.

\section{The quantum double model}
\begin{figure}
\includegraphics[angle=0, width=0.5\textwidth]{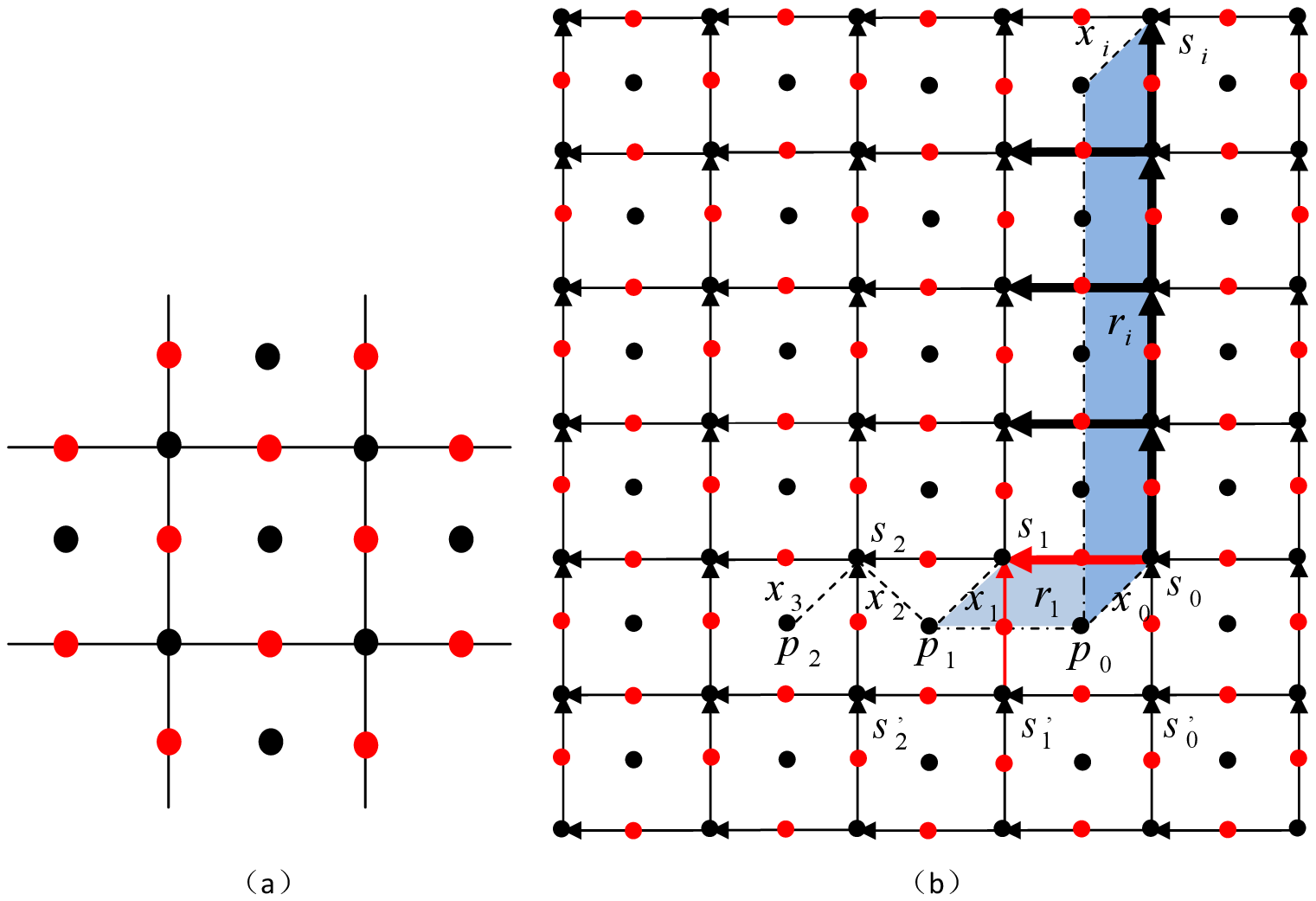}
\caption{\footnotesize{(Color online)(a) A spin lattice for the quantum double model. The system particles
 (red) reside on the edges of the lattice.
 All particles on edges that meet at a vertex interact.
 Ancillary particles (black) reside on the vertices and center of the faces. They can interact with their nearest system
 particles. (b) Short and long ribbons on the lattice indicated by the blue areas. Arrows label the orientation of the edges.
 The short ribbon
 operator for ribbon $r_{1}$ connects adjacent sites $x_{1}$ and $x_{0}$ and acts on the two red edges.
 The long ribbon $r_{i}$
 connects faraway sites $x_{i}$ (dashed) and the base site $x_{0}$. The corresponding ribbon operator acts on the edges in thick lines.
 }}
\label{fig:lattice}
\end{figure}

 Kitaev's quantum double model is a spin Hamiltonian which is a sum of
quasi-local operators in a two-dimensional lattice \cite{A.K1}. Its ground
states are invariant under gauge transformations generated by some finite group.
In this work, we consider the non-Abelian group $G=S_{3}$. For the convenience of
discussion and without loss of generality, we focus on a square lattice as shown
in Fig. \ref{fig:lattice}. Particles live on the edges of the lattice, and their
internal Hilbert space is described by the group $G$,
$\mathcal{H}=\{|g\rangle:g\in G\}$. In Fig. \ref{fig:lattice}, arrows are used to label the
orientation of the edges in the lattice. Reversing the direction of a particular
arrow is equivalent to making the basis change $|z\rangle\rightarrow|z^{-1}\rangle$
(where $|z\rangle\in\mathcal{H},z\in G$)for the corresponding qudit.
To describe the model, one need to define four types of
linear operators acting on the
Hilbert space $\mathcal{H}$: $L_{\pm}^{g}, g\in G$ and $T_{\pm}^{h}, h\in G$.
Their effect on the basis state $|z\rangle$ is as follows:
\begin{equation} \label{eq:LT}
L_{+}^{g}|z\rangle=|gz\rangle, L_{-}^{g}|z\rangle=|zg^{-1}\rangle,
\end{equation}
\begin{equation}
T_{+}^{h}|z\rangle=\delta_{h,z}|z\rangle,T_{-}^{h}|z\rangle=\delta_{h^{-1},z}|z\rangle\ .
\end{equation}

To study the effect of these operators on individual qudits, we use $j$ and $s$ to denote
an edge of the lattice and one of its endpoints. Then one can define an operator $L_{g}(j,s)$ as
follows: if $s$ is the tail of the arrow on edge $j$ then $L_{g}(j,s)$ is $L_{-}^{g}$
acting on the $j$-th particle. If $s$ is the head of the arrow on edge $j$, $L_{g}(j,s)$
is $L_{+}^{g}$ acting on the same particle. Similarly, if $p$ is the left (right)
adjacent face of edge $j$ then $T_{h}(j,p)$ is $T_{-}^{h}$ ($T_{+}^{h}$) acting on the
$j$-th particle. Local gauge transformations and magnetic charge operators for a vertex
$s$ and its adjacent face $p$ are defined as
\begin{equation} \label{eq:GT1}
A_{g}(s,p)=A_{g}(s)=\prod_{j\in star(s)}L_{g}(j,s)
\end{equation}
and
\begin{equation} \label{eq:GT2}
B_{h}(s,p)=\sum_{h_{1}\cdots h_{k}=h}\prod_{m=1}^{k}T_{h_{m}}(j_{m},p).
\end{equation}
Here, $j_{1}\cdots j_{k}$ are the boundary edges of face $p$, starting from and ending at
vertex $s$ and enumerated in the counterclockwise direction. Notice that $B_{h}(s,p)$ is a
projector into
states with magnetic flux $h$ on face $p$.

The quantum double model Hamiltonian defined by Kitaev is
\begin{equation} \label{eq:H}
H=\sum_{s}(1-A(s))+\sum_{p}(1-B(p)),
\end{equation}
where
\begin{equation} \label{eq:GT}
A(s)=\frac{1}{|G|}\sum_{g}A_{g}(s), B(p)=B_{e}(s,p),
\end{equation}
$e$ is the identity
element of the group. The ground states satisfy $A(s)|GS\rangle=|GS\rangle$,
$B(p)|GS\rangle=|GS\rangle$ for all $s$ and $p$, and exited states involve some
violations of these conditions. Because the projection operators $A(s)$ and $B(p)$
are localized, excitations are particle-like living on vertices or faces, or
both, where the ground state conditions are violated. A combination of
a vertex and
an adjacent face will be called a \textit{site} (see Fig. \ref{fig:lattice}).

A detailed examination of the excitation properties in the quantum double model is
presented in \cite{A.K1}. Here, we just give a brief description sufficient for our purpose.
The quantum double of group $G$, $D(G)$, is a quasi-triangular Hopf
algebra that has a set of linear bases $D_{h,g}(x)=B_{h}(x)A_{g}(x)$ on site $x=(s,p)$.
Though $D_{h,g}(x)$ is defined on each site, the structure of the quantum double
does not depend on the \textit{specific site}. Quasi-particle excitations in this system
can be created by ribbon operators $F^{g,h}(r)$ introduced by Kitaev~\cite{A.K1}
which we will discuss in detail later. For a system with $n$ quasi-particles, one can use
$\mathcal{L}[n]$ to denote the quasi-particles' Hilbert space. By investigating how
local operators
$D_{h,g}(x)=B_{h}(x)A_{g}(x)$ act on this Hilbert space, one can define types and
subtypes of these quasi-particles according to their internal states.
It turns out that the types of
the quasi-particles have a one-to-one correspondence with the irreducible representations
of $D(G)$. These representations are labeled by $\Pi_{R(N_{[\mu]})}^{[\mu]}$,
where $[\mu]$ denotes a conjugacy class of $G$
which labels the magnetic charge. $R(N_{[\mu]})$ denotes a unitary irrep
of the centralizer of an arbitrary element
in the conjugacy $[\mu]$ and it labels the electric charge. The 8 irreps for $D(S_3)$ are listed in the Appendix.B.
Once the types of the
quasi-particles are determined they never change. Besides the type, every
quasi-particle has a local degree of freedom, the subtype. The Hilbert space of
$n$ quasi-particles then splits according to
$\mathcal{L}[n]=\bigoplus_{d_{1}\cdots d_{n}}\mathcal{K}_{d_{1}}\bigotimes\cdots\bigotimes\mathcal{K}_{d_{n}}\bigotimes\mathcal{M}_{d_{1}\cdots d_{n}}$,
where $d_{m}$ is the type of the $m$th quasi-particle, and $\mathcal{K}_{d_{m}}$ is the space
of local degrees of freedom of the $m$-th quasi-particle. $\mathcal{K}_{d_{m}}$ is just the representation
spaces of $(\Pi_{R(N_{[\mu]})}^{[\mu]})_{m}$ and it is spanned by the basis
$\{|(\nu_{(L)},\xi_{(L)})\rangle=|\nu_{(L)}\rangle\otimes|\xi_{(L)}\rangle\}$, where $\nu_{(L)}\in [\mu]_{m}$ and $|\xi_{(L)}\rangle$ is the
basis of the representation space of
$R(N_{[\mu]})_{m}$. $\mathcal{M}_{d_{1}\cdots d_{n}}$ is the topologically protected
space. It is inaccessible by local measurements and insensitive to local
perturbations. When we braid these quasi-particles, the protected space undergoes
some unitary transformation, but the type and the subtype of the quasi-particles do
not change under braiding.

Unfortunately, though $\mathcal{M}_{d_{1}\cdots d_{n}}$ is
topologically protected, it dose not have a tensor product
structure. This makes it difficult to use it directly for quantum
computation. However, if we choose a \textit{base site} and
connect it with other sites by non-intersecting ribbons (see Fig.
\ref{fig ribbon} (a)), we can create quasi-particle excitations
whose associated protected space can be described with a tensor
product structure. Each quasi-particle excitation provides a
protected subspace, and for a quasi-particle whose type is
$\Pi_{R(N_{[\mu]})}^{[\mu]}$, the corresponding protected subspace
is spanned by
$\{|(\nu,\xi)\rangle=|\nu\rangle\otimes|\xi\rangle\}$, where
$\nu\in [\mu]_{m}$ and $|\xi\rangle$ is the basis of the
representation space of $R(N_{[\mu]})_{m}$. The overall
topologically protected space is the tensor product of these
protected subspaces. These quasi-particle excitations are
non-Abelian anyons. Their braiding and fusion rules are given
in~\cite{A.K1,Wild}, and a brief summary of $S_3$ group and
$D(S_3)$ anyons is given in the Appendix. If the conjugacy class
$[\mu]=[e]$, where $e$ is the identity group element, the
quasi-particle is a pure electric excitation, we can simply use
$|\xi\rangle=|(\nu,\xi)\rangle$ to denote its topological state.
If the unitary irrep $R$ is the identity representation, the
quasi-particle is a pure magnetic charge excitation, and we use
$|\nu\rangle=|(\nu,\xi)\rangle$ to denote its topological state.
For these pure charge excitations, their local degrees of freedom
can also be simplified in the same way. These pure charge
excitations based on $D(S_3)$ are sufficient for universal TQC
\cite{Mochon}. The braiding rules for pure charge excitations are
\begin{equation} \label{eq:braid1}
\mathcal{R}|\nu_{1}\rangle|\nu_{2}\rangle=|\nu_{1}\nu_{2}\nu_{1}^{-1}\rangle|\nu_{1}\rangle,
\end{equation}
and
\begin{equation} \label{eq:braid2}
\mathcal{R}^{2}|\nu\rangle|\xi_{n}^{R}\rangle=|\nu\rangle|R(\nu)_{mn}\xi_{m}^{R}\rangle.
\end{equation}
In the above equations, we have neglected the local degree of freedom. $R$ is the unitary
irrep of $S_{3}$ corresponding to pure electric charge excitation. $\nu$ corresponds
to pure magnetic charge, and $\mathcal{R}$ denotes the counterclockwise exchange of the two
anyonic excitations. The exchange between pure electric charges is trivial.

\section{Simulation of the quantum double model}

The quantum double Hamiltonian in Eq. (\ref{eq:H}) is very difficult to realize because it requires interactions between
more than two particles. Instead of engineering a Hamiltonian with genuine multiparticle
interactions, we take a dynamic approach to construct appropriate ribbon operators using local operations
involving no more than two qudits at a time.

We focus on the square lattice model shown in Fig. \ref{fig:lattice}. System qudits live on the edges.
In previous schemes \cite{ABVC1,ABVC2}
based on trapped atoms in 2d optical lattices, ancillary qudits
are also used. They are not
needed if the system qudits are addressable, as is the case
in a solid-state system. Nevertheless, in order to demonstrate the wide applicability of our scheme, we will present
it in a system with ancillary qudits that reside on the vertices and center of faces too
(see Fig. \ref{fig:lattice} (a)).
The ancillary qudits are assumed to be addressable.

To simulate the quantum double model, we apply single-qudit gates on
individual ancillary qudits, as well as single-qudit gates on all
system qudits simultaneously (since they are not addressable).
In addition, we need the two-qudit diagonal
phase gate $U=\exp(i\phi|g_{i}\rangle_{A}\langle g_{i}|\otimes|g_{i}\rangle_{B}\langle g_{i}|)$ between
an ancillary qudit and its adjacent system qudits. With these operations, all two-qudit controlled rotations
between the ancilla and system qudits can be constructed efficiently \cite{BOB,ABVC2}.

\subsection{Ground state preparation}
We first need to prepare the system in the ground state, and this can be achieved using a similar method as in
\cite{ABVC1,ABVC2}. Initially, all
system qudits and face ancillary qudits are in state $|e\rangle$ and vertex ancillary qudits are in
state $\frac{1}{\sqrt{|G|}}\sum_{g}|g\rangle$. Note that in this initial state, $B(p)=1$ for all $p$. To create the ground state, we only need to make the symmetrized gauge transformation
projection $A(s)$ Eq. (\ref{eq:GT}) on every vertex.
This symmetrized gauge transformation is carried out from left to right and top to bottom on all vertices in the lattice.
The ground state is created as a result.
Now we take $s_0$ in Fig. \ref{fig:lattice} (b) as an example and show how to make the projection $A(s_0)$. Projections $A(s)$
on other vertices are similar. Using the vertex ancillary
qudit $s_0$ as the controlled qudit, we apply on system qudits the controlled gauge transformation
$\sum_{g}|g\rangle_{s_0}\langle g|\otimes A_{g}(s_0)$. After this operation, the state of system qudit on edge $[s_0,s_0']$
(i.e. the bottom edge of $s_0$ in Fig. \ref{fig:lattice} (b)) is the same as state of ancilla $s_0$. Then, disentangle the ancillary qudits from
the system by controlled two-qudit gate $\sum_g|g\rangle_{[s_0,s_0']}\langle g|\otimes L_+^{g^{-1}}(s_0)$ between
the system qudit on edge $[s_0,s_0']$ and ancilla $s_0$. Now the ancilla $s_0$ is in state $|e\rangle$, and we make
the operation $A(s_0)=\frac{1}{\sqrt{|G|}}\sum_g A_g(s_0)$.

Alternatively, one can realize the above disentanglement step by first making proper measurements on the vertex
ancilla, and then performing appropriate corrections on the system qudits according to the measurement outcome \cite{ABVC2}.
For the correction, one needs single qudit unitary
operations on system qudits. Assuming one of the operations on single system qudit is $U$ , without addressability of system qudits, we can still realize $U$
by preparing
an adjacent (and addressable) ancillary qudit $a$ in the state $|h\rangle$ and performing the controlled unitary operation
$|h\rangle_a\langle h|\otimes U+1_{|G|-1}\otimes 1_{|G|}$.
Similarly, all single-qudit operations on system qudits in our following schemes can be
implemented by this technique.

\subsection{Anyon creation and braiding}

The central issue in our simulation is anyon creation from ground state. In order to create a topologically protected space with a tensor
structure that can be used for TQC, we
choose an arbitrary site $x_{0}$ as our base site as shown in Fig. \ref{fig:lattice} (b). The key then is to find
ways to create anyons by
dynamically performing the
ribbon operators $F^{g,h}(r)$ connected to $x_0$ on the ground state (see Fig. \ref{fig:lattice} (b)).
This will allow us to create arbitrary topological states of a given type by applying the superposition
ribbon operator $\sum_{z\in S_3}\alpha_zF^{\mu,z}(r)$, where $\mu$ corresponds to the anyon type and
$\alpha_z$ are the coefficients.

These ribbon operators $F^{g,h}(r)$ act as follows \cite{A.K1}
{
\unitlength=14mm
\def\relvshift{0.5}
\def\hsign#1{\vbox{\hbox to \unitlength {\hfil#1\hfil} \kern 0.5pt}}
\newcount\abc
\def\leftarrow(#1,#2)#3{\abc=#1 \advance\abc 1
     \put(\abc,#2){\vector(-1,0){1}} \put(#1,#2){\hsign{{\tiny $#3$}}}}
\def\uparrow(#1,#2)#3{\abc=#2 \advance\abc -1
     \put(#1,\abc){\vector(0,1){1}} \put(#1,\relvshift){\kern 0.2pt {\tiny $#3$}}}

\begin{equation} \label{ribop}
  \begin{array}{rl}
    F^{(h,g)}(r)\ &
    \raisebox{-\relvshift\unitlength}{
      \begin{picture}(3,1.2)
        \leftarrow(0,1){y_1} \leftarrow(1,1){y_2} \leftarrow(2,1){y_3}
        \uparrow(0,1){y_1'} \uparrow(1,1){y_2'} \uparrow(2,1){y_3'}
      \end{picture}
    }
    \qquad  \bigskip \\
    =    \quad    \delta_{g,\,y_1y_2y_3}\ &
    \raisebox{-\relvshift\unitlength}{
      \begin{picture}(3.3,1.2)
        \leftarrow(0,1){y_1} \leftarrow(1,1){y_2} \leftarrow(2,1){y_3}
        \uparrow(0,1){hy_1'}
        \uparrow(1,1){y_1^{-1}hy_1\,y_2'}
        \uparrow(2,1){(y_1y_2)^{-1}h(y_1y_2)\,y_3'}
      \end{picture}
    }
  \end{array}
\end{equation}
}
The ribbon operators $F^{(h,g)}(r)$ commute with every projector $A(s)$ and $B(p)$, except when $(s,p)$
is on either end of the ribbon. Therefore, the ribbon operator creates excitations on
both ends of the ribbon. To simplify our discussion, we will use long ribbon operators to create excitations
at sites (infinitely) far
away from the base site $x_0$. When one examines the excitations on these sites, the base site
$x_0$ is at an (infinitely) faraway location, and one can ignore the effect of the excitations on the base
site $x_0$  \cite{A.K1}.
By doing so, when we apply a long ribbon operator connected to $x_0$
we can solely focus on the quasi-particle excitation on the other end of the ribbon \cite{A.K1}.

In the spirit of this consideration, we will first propose a method to realize short ribbon operators to
create anyons near the base site $x_0$, and then show how to move them faraway from $x_0$ to implement a long ribbon operator.
This protocol to move anyons around is  needed for anyon braiding and
measurement too, and thus is critical ingredient in our scheme.

Now let us study how to perform a short ribbon operator in the general superposition form
\[\sum_{z\in S_3}\alpha_zF^{\mu,z}(r_1),\]
where $r_1$ is a ribbon connecting the base site $x_{0}$ and a nearby site $x_{1}$
in Fig. \ref{fig:lattice} (b). This allows us to create an anyon at $x_{1}$ in an arbitrary
topological state with a certain type determined by $\mu$.

In order to perform the ribbon operator $\sum_{z\in S_3}\alpha_zF^{\mu,z}(r_1)$, we need the projection operation
$\sum_i\alpha_{z_i}|z_i\rangle\langle z_i|$($z_i\in\{z\in G|\alpha_{z}\neq0\},i=1,2,\cdots,m$) on system qudit
$[s_{0},s_{1}]$ and single qudit gate $L_+^{\mu}$ on system qudit $[s_{1},s_{1}']$. Using a single projection
measurement that corresponds to operation $|g\rangle\langle g|$, together with
appropriate gauge transformations,
we can realize the projection $\sum_i\alpha_{z_i}|z_i\rangle\langle z_i|$:

\[
\begin{split}
&\sum_i\alpha_{z_i}A_{z_i}(s_1)A_{g_1^{-1}}(s_1)|g_1\rangle_{[s_0,s_1]}\langle g_1||GS\rangle \\
=&\sum_i\alpha_{z_i}A_{z_i}(s_1)A_{g_1^{-1}}(s_1)|g_1\rangle_{[s_0,s_1]}\langle g_1|A_{g_1}|GS\rangle \\
=&\sum_i\alpha_{z_i}A_{z_i}(s_1)|e\rangle_{[s_0,s_1]}\langle e|A_{z_i^{-1}}(s_1)|GS\rangle \\
=&\sum_i\alpha_{z_i}|z_i\rangle_{[s_0,s_1]}\langle z_i||GS\rangle.
\end{split}
\]
We have used $A_{g}(s_1)\equiv 1$ as there is no excitation at vertex $s_1$(i.e. $A(s_1)=1$).
Following this idea, we can design the following protocol to
perform the ribbon operator $\sum_{z\in S_3}\alpha_zF^{\mu,z}(r_1)$.

\begin{enumerate}
 \item Measure system qudit on edge $[s_{0},s_{1}]$ in the basis $\{|g\rangle\}$. This is done by applying
  the two qudit unitary between system qudit on edge $[s_{0},s_{1}]$ and ancillary qudit $p_{0}$,
  $\sum_{g}|g\rangle_{[s_{0},s_{1}]}\langle g|\otimes L_{+}^{g}(p_{0})$, and then measuring ancilla $p_{0}$ in the
  basis $\{|g\rangle\}$. If the outcome is $|g_{1}\rangle$,
  perform gauge transformation $A_{g_{1}}^{-1}(s_{1})$ (This step is equivalent to a projection $|e\rangle\langle e|$ on edge $[s_{0},s_{1}]$).

 \item Prepare the ancillary qudit $s_{1}$ in the state
  $|0_{[\alpha_z]}\rangle_{s_{1}}\propto\sum_{i}\alpha_{z_i}|z_i\rangle$
  and apply the controlled gauge transformation $\sum_{g}|g\rangle_{s_{1}}\langle g|\otimes A_{g}(s_{1})$.

 \item Measure the ancillary qudit $s_1$ in the basis $\{|k_{[z]}\rangle=Z_{[z]}^{k}|0_{[z]}\rangle\}$, where
  $Z_{[z]}^{k}=\sum_{z_{j}}\exp(i2\pi kj/m)|z_{j}\rangle\langle z_{j}|$ and $|0_{[z]}\rangle=\frac{1}{\sqrt{m}}\sum_{i}|z_{i}\rangle$. For outcome $k_{[z]}$, apply the single
  qudit operation $Z_{[z]}^{k}$ on the system qudit on edge $[s_{0},s_{1}]$ as a correction. Alternatively, this
  disentanglement step can be done by applying two-qudit gate
  $\sum_{g}|g\rangle_{[s_{0},s_{1}]}\langle g|\otimes L_{+}^{g^{-1}}(s_{1})$ between system qudit on edge $[s_{0},s_{1}]$
  and ancilla $s_1$.

 \item Apply operation $L_{\mu}(j,s_{1})$, where $j$ is the edge $[s_{1},s_{1}']$.
\end{enumerate}

By these steps, we can create an anyon at $x_{1}$ with arbitrary type and topological state.

We first take the pure
magnetic charge excitation
\[|\mu_{(L)};\nu\rangle=|C|^{\frac{1}{2}}\sum_{z\in G:z^{-1}\mu_{(L)}z=\nu}F^{\mu_{(L)},z}(r_{1})|GS\rangle\]
as an example. Here, $\mu_{(L)}$ is an element of the conjugacy class $C$. It corresponds to the magnetic flux at
site $x_{1}$
and characterizes the local degree of freedom. $\nu$, on the other hand, corresponds to the magnetic flux across
a counterclockwise circle, starting and ending at the base site $x_{0}$ and surrounding only the quasi-particle
at $x_{1}$.
It is the topological state that we are interested in. Notice that, for this pure
magnetic charge anyon
$|\mu_{(L)};\nu\rangle$, the corresponding coefficients $\alpha_{z_i}\neq0$ only when $z_i\in\{z\in G|z^{-1}\mu z=\nu\}$,
and $\alpha_{z_i}=|C|^{\frac{1}{2}}$ for all $z_i\in\{z\in G|z^{-1}\mu z=\nu\}$. So,
in step 2 given above, we prepare the ancillary
qudit at $s_{1}$ in the state $|0_{[z]}\rangle_{s_{1}}\propto\sum_{i}|z_i\rangle$, with $z_i\in\{z\in G|z^{-1}\mu z=\nu\}$.
As another example, we look at the pure electric charge excitation
$|\xi_{(L)};\eta\rangle=|R|^{\frac{1}{2}}\sum_{g_{i}}R_{\xi_{(L)}\eta}(g_{i})F^{e,g_{i}}(r_{1})|GS\rangle$, where $R$ is a irreducible
representation of the group $G$(see the appendix), $\xi_{(L)}$ is the local degree of freedom, and $\eta$ characterizes the topological state.
To create this state,
we need to prepare the ancillary
qudit at $s_{1}$ in the state $|0_{[R_{\xi_{(L)}\eta}]}\rangle_{s_{1}}\propto\sum_{g_{i}}R_{\xi_{(L)}\eta}(g_{i})|g_{i}\rangle$
in step 2. As a more generic example, we look at the
dyonic combination excitation $|(\nu,\xi)\rangle$ where $\nu$ and $\xi$ are both topological degrees of freedom
(local degrees of freedom ignored). To create this state, we perform corresponding
superpositions of ribbon operators $\sum_{h\in S_3}\alpha_hF^{g,h}(r)$ with a fixed $g\in[\nu]$, and appropriate
coefficients $\alpha_h$
determined by the topological state.

Once we created anyonic excitations on sites close to the base site $x_0$, we need to move them far away from
$x_0$ and also
braid them. For these purposes, we need two basic movements. We need to move the excitation from the original
site to an adjacent site sharing a common face, and from the original site to an adjacent site sharing a common vertex.
In Fig. \ref{fig:lattice} (b), these correspond to moving the excitation from site $x_1$ to $x_2$, and from $x_2$ to
$x_3$.

For the first quasi-particle movement from site $x_{1}$ to
site $x_{2}$, mathematically, this means mapping the group element corresponding to state for qudit $[s_0,s_1]$ to a product
of group elements corresponding to states for
qudits $[s_1,s_2]$ and $[s_0,s_1]$ (i.e. coherently mapping from
$|g_1\rangle_{[s_0,s_1]}|g_2\rangle_{[s_1,s_2]}|\psi\rangle_{rest}$ to $ |g_1'\rangle_{[s_0,s_1]}|g_2'\rangle_{[s_1,s_2]}|\psi'\rangle_{rest}$
with $g_2'g_1'=g_1$ and $|\psi\rangle_{rest}$ referring to the state of the qudits in the rest of the system)
and moving the flux at $x_{1}$ to $x_{2}$. To do this,
we first perform the
projection operation $|e\rangle\langle e|$ on
the qudit on edge $[s_{1},s_{2}]$(see the first step of anyon creation).
Now qudits
$[s_1,s_2]$ and $[s_0,s_1]$ are in state
$|g_2'\rangle_{[s_1,s_2]}|g_1'\rangle_{[s_0,s_1]}$ with $g_2'=e$ and $g_1'=g_1$, and we have $g_2'g_1'=g_1$.
The flux at site $x_{1}$ doesn't change and now the flux at site $x_2$
is the same as site $x_{1}$. Then, recall the action of the ribbon operator, we need to erase redundant
excitation at site $x_{1}$ to finish this movement.
This can be done by applying the symmetrized gauge
transformation
$A(s_{1})$ in Eq. \eqref{eq:GT} at vertex $s_{1}$(for operation $A(s_1)$, see the ground state preparation).

The second basic movement
from site $x_{2}$ to $x_{3}$ has been studied in
\cite{ABVC2}.
Here we give a similar scheme with a simpler disentanglement step.
We first coherently map the flux at site $x_{2}$ to the ancillary qudit at $p_{1}$ by the controlled operation
\[\Lambda(x_2,p_1)=\sum_{g\in S_3}B_g(x_2)\otimes L_+^g(p_1).\] Then we apply the controlled unitary
$\sum_{h\in S_3}|h\rangle_{p_1}\langle h|\otimes L_+^{h^{-1}}([s_{2},s_{2}'])$ to move the flux from
site $x_2$ to site $x_3$, just as in \cite{ABVC2}. Finally we disentangle the ancillary qudit $p_1$ from the system by first swapping
ancilla $p_1$ and $p_2$ and then applying $\Lambda(x_3,p_2)^{-1}$. The controlled operation $\Lambda(x,p)$ can
be decomposed into elementary two-qudit controlled rotation operators with each edge surrounding $p$ as
a control and the ancilla as the target \cite{ABVC2}.

The combination of these two basic movements allows us to move any anyonic excitations
around. By applying them repeatedly, we can realize a long ribbon and move the
quasi-particle to an arbitrary site faraway from the base site $x_{0}$ and also braid the anyons.
Notice that, after we create an anyonic excitation at a location faraway from the base site
$x_0$, we can use the same procedure to create more anyonic excitations by first applying a
short ribbon operator and then moving the excitation away. In this process, since the
excitations created earlier have been moved away from around the base site, they will not
affect the short ribbon operations later.

For TQC based on this quantum double model $D(S_3)$,
two kinds of ancillary vacuum states are needed,
the chargeless pair state of pure
electric charges $|I_{|R|}\rangle=\frac{1}{\sqrt{|R|}}\sum_\eta|\eta\rangle_{R}\otimes|\eta\rangle_{R^{*}}$ and the
chargeless pair state of pure magnetic charges
$|I_{[\mu]}\rangle=\frac{1}{\sqrt{|[\mu]|}}\sum_{\nu\in[\mu]}|\nu\rangle\otimes|\nu^{-1}\rangle$~\cite{Mochon}.
Here, $\eta$ and $\nu$
are topological states and we omit the local degrees of freedom, and $R$ is a irreducible
representation of the group $S_3$(see the appendix). The pair excitations localized at
sites $x_{i}$ and $x_{j}$(Fig. \ref{fig ribbon} (a)) can be created by a ribbon operator connecting $x_{i}$ and
$x_{j}$. If a pair of anyons is created from vacuum, the total topological charge of them is zero automatically. So
we have $|I_{|R|}\rangle=|R|^{\frac{1}{2}}\sum_{z}R_{\xi_{1(L)}\xi_{2(L)}}(z)F^{e,z}(r_{i,j})|GS\rangle$, and
$|I_{[\mu]}\rangle=|C|^{\frac{1}{2}}\sum_{z:z^{-1}\mu_{1(L)} z=\mu_{2(L)}}F^{\mu_{1(L)},z}(r_{i,j})|GS\rangle$, where $r_{i,j}$
is a ribbon that connects $x_{i}$ and $x_{j}$, and $\xi_{1(L)}$, $\xi_{2(L)}$, $\mu_{1(L)}$ and $\mu_{2(L)}$ are all local degrees of freedom.
Obviously, we can dynamically create this excitation pair by performing the corresponding ribbon operators using the
method discussed earlier.
Therefore, we can simulate universal TQC in principle.

\subsection{Fusion and Topological state Measurements}
We must be able to measure the anyon states in order to simulate the quantum double model. Measurement is also needed for TQC.
An indirect but technically less challenging method is to perform a fusion
measurement. It probes the fusion rules of the anyons by which we can confirm the topological nature of the system.
If we want to study the braiding rules, then a direct
measurement of anyons' topological states is necessary. These measurements can be accomplished by closed ribbon operator
projection and interference.

There are a few ways to do a fusion measurement. One can first fuse two anyons to one and then measure the type of
the new anyon. Alternatively, one can move two anyons close and perform appropriate closed ribbon operators
that encircle them, which does not destroy the two anyons since they are not really fused. In our dynamical simulation,
this is the realistic approach since we can only move two anyons close rather than fuse them.

\begin{figure}
\includegraphics[angle=0, width=0.5\textwidth]{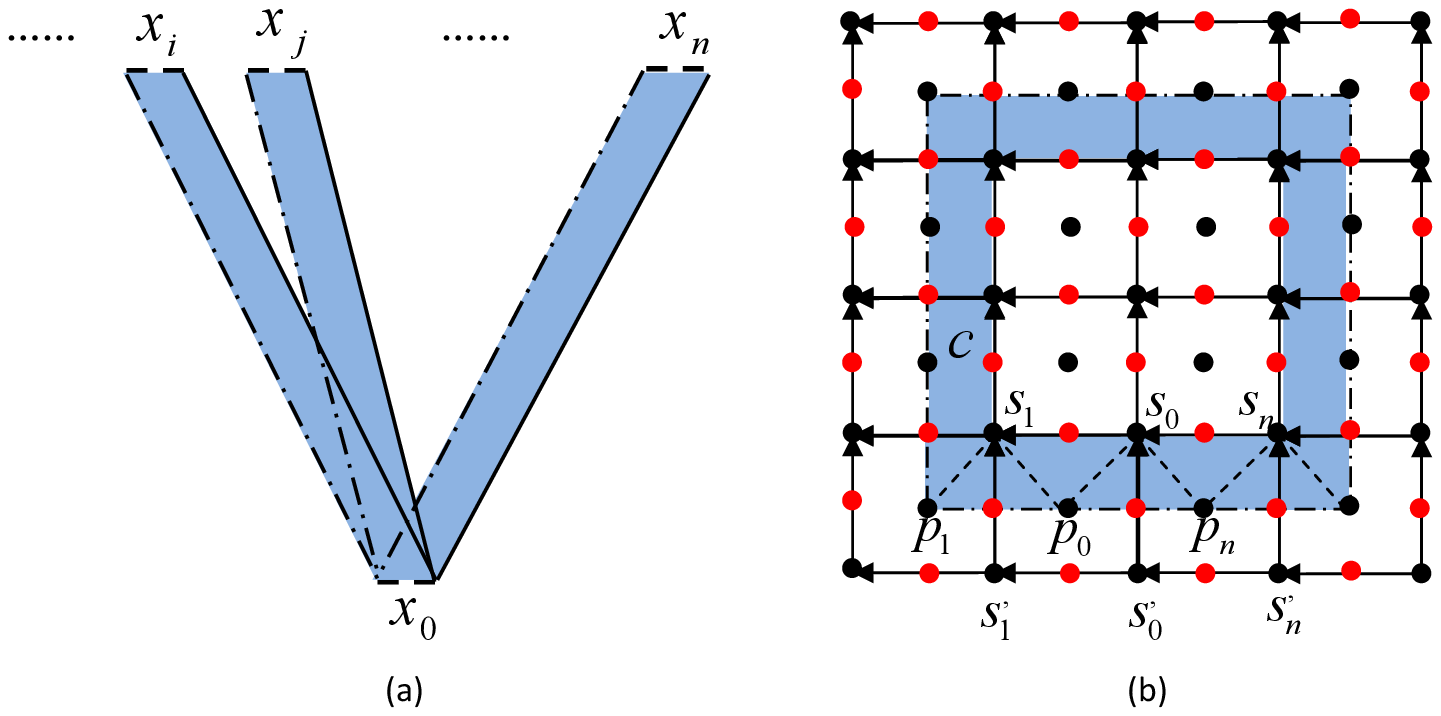}
\caption{\footnotesize{(a)Non-intersecting ribbons with one common end $x_0$(the base site).
(b)A closed ribbon for fusion measurement.}}
\label{fig ribbon}
\end{figure}

For TQC,  the only measurement we need is to detect whether there is a quasi-particle left or whether two anyons
have vacuum quantum numbers when they fuse ~\cite{Mochon}. Assuming the closed ribbon
$c$ in Fig. \ref{fig ribbon} (b) encircles only the two anyons to be fused, this measurement corresponds to the projection ribbon operator \cite{BM}
\begin{equation} \label{eq:FM}
\frac{1}{|G|}\sum_{g\in G}F^{g,e}(c)
\end{equation}
which casts the two anyons into the vacuum quantum number state when they fuse.
We then focus on this projection ribbon operator and show how to realize it. In principle, projection operators corresponding
to other fusion channels can be realized in a similar way.

As shown in Fig. \ref{fig ribbon} (b), to make the projection in Eq.\eqref{eq:FM}, all ancillary qudits are initially
prepared in state $|e\rangle$. We first prepare ancillary qudit
$p_0$ in state
$|0_{[G]}\rangle=\frac{1}{\sqrt{|G|}}\sum_{g}|g\rangle$.
And if we apply the controlled ribbon operator $\sum_{g\in G}|g\rangle_{p_0}\langle g|\otimes F^{g,e}(c)$, we get the final
state of the system and ancilla $p_0$: $|\psi_f\rangle\propto\sum_{g\in G}|g\rangle_{p_0} F^{g,e}(c)|\psi_i\rangle$.
We can then measure $p_0$ in basis $\{|k_{[G]}\rangle=Z_{[G]}^{k}|0_{[G]}\rangle\}$, where
$Z_{[G]}^{k}=\sum_{g_{j}}\exp(i2\pi kj/|G|)|g_{j}\rangle\langle g_{j}|$. If the outcome is $|0_{[G]}\rangle$, the
projection in Eq. \eqref{eq:FM} succeeds, and the two anyons in
ribbon $c$ have vacuum quantum number when they fuse.
If the outcome is not $|0_{[G]}\rangle$, the projection fails, and there is a quasi-particle left. The controlled closed ribbon operator can be realized
by first performing a short controlled ribbon operator and then coherently moving one end of the ribbon along $c$ until the ribbon
is closed. Notice that $F^{g,e}(c)$ contains projections.
And for the last two step movements that close the ribbon, there may be excitations at the target site of the movements,
but our movement scheme above only works when there are no excitations at the target sites. So
we must carefully close the ribbon. The following steps should be followed.

\begin{enumerate}

  \item Prepare ancillary qudit $p_0$ in state $|0_{[G]}=\frac{1}{\sqrt{|G|}}\sum_{g}|g\rangle$.
  Perform the projection operation $|e\rangle\langle e|$ on the qudit on edge $[s_{0},s_{1}]$ (see the
  first step of anyon creation).

  \item Apply the controlled two-qudit gate $\sum_{g\in G}|g\rangle_{p_{0}}\langle g|\otimes L_{g}(s_{1},[s_{1},s_{1}'])$
  between the ancillary qudit
  $p_{0}$ and system qudit $[s_{1},s_{1}']$.

  \item Move the excitation at $(s_{1},p_{1})$ along the ribbon $c$ in the clockwise direction to site
  $(s_{n},p_{n})$ in Fig. \ref{fig ribbon} (b).

  \item Measure qudit $[s_{n},s_{0}]$ in basis $\{|g\rangle:g\in G\}$. If the outcome is not $|e\rangle$,
  projection Eq.(\ref{eq:FM}) fails and there is a quasi-particle left when the two anyons fuse. If the outcome is $|e\rangle$, perform
  $A(s_{n})$ in Eq. \eqref{eq:GT}.

  \item Apply the controlled two-qudit gate
  $\sum_{g\in G}|g\rangle_{p_{0}}\langle g|\otimes L_{g}(s_{0},[s_{0},s_{0}'])$ between the ancillary qudit $p_{0}$ and
  system qudit $[s_{0},s_{0}']$. Measure ancillary qudit $p_{0}$ in basis ${|k_{[G]}\rangle=Z_{[G]}^{k}|0_{[G]}\rangle}$,
  where $Z_{[G]}^{k}=\sum_{g_{j}}\exp(i2\pi kj/|G|)|g_{j}\rangle\langle g_{j}|$. If the outcome is $|0_{[G]}\rangle$,
  the projection is done and the two anyons in ribbon $c$ have vacuum quantum number when they fuse. Otherwise, there is
  a quasi-particle left behind.

\end{enumerate}

Aside from fusion measurements, we can also measure the topological states of the anyons directly in an
interference experiment. This can be accomplished by using controlled braiding. In Fig. \ref{fig:lattice} (b),
assume we want to measure the anyon at site $x_{i}$ which is in some unknown topological state $|x\rangle$.
We make use of the ancillary anyon at
site $x_{1}$ Fig. \ref{fig:lattice} (b) (assuming $x_{1}$ is faraway from the base site $x_{0}$) which is prepared in the topological state $|a\rangle$.
To perform the interference measurements, initially, we prepare all ancillary qudits in state $|e\rangle$. Then
\begin{enumerate}
 \item prepare the ancillary qudit $p_{1}$ in state $\frac{1}{\sqrt{2}}(|e\rangle+|h\rangle)$;
 \item coherently braid the anyon
  at $x_{1}$ around $x_{i}$ counterclockwise only when the ancillary particle $p_1$ is in state $|h\rangle$;
 \item measure the ancilla $p_{1}$ in basis $\{\frac{1}{\sqrt{2}}(|e\rangle\pm|h\rangle)\}$.

\end{enumerate}
The outcome of the measurement on the ancilla $p_{1}$ is $m=\pm1$. The probability distribution for either result is
\begin{equation} \label{eq:interfer}
P(m=1)-P(m=-1)=Re[\langle a|\langle x| \mathcal{R}^{2}|x\rangle|a\rangle]
\end{equation}
Therefore, changing the state $|a\rangle$ and measuring the probability distributions allows us to obtain the real
part of the interference amplitudes.
Similarly, measuring the ancilla $p_{1}$ in the basis $\{\frac{1}{\sqrt{2}}(|e\rangle\pm i|h\rangle)\}$ yields the imaginary part of the interference
amplitudes. With these interference amplitudes, one can determine the topological state $|x\rangle$ \cite{Wild}.

The most critical step for the interference measurement is the controlled braiding. It can be
realized by controlled
movements of anyons and braiding operations.
First, we coherently move the anyon at $x_{1}$
to $x_{3}$ if and only if the ancilla $p_{1}$ is in state $|h\rangle$. Then we braid anyon at $x_{3}$ around $x_{i}$.
Finally, we perform a controlled movement of the anyon from $x_{3}$ back to $x_{1}$.

For the controlled movement from $x_{1}$
to $x_{3}$ and back, we first perform the projection operation $|e\rangle\langle e|$ on
the system qudit $[s_{1},s_{2}]$. Now the product of group elements corresponding to states of qudits $[s_0,s_1]$ and $[s_1,s_2]$
is the same as group element corresponding to state of
qudit $[s_0,s_1]$, and the flux at site $x_{2}$ is the same as site $x_{1}$. Recall the action of ribbon operators, we can make a controlled erase of the quasi-excitations at $x_2$ and $x_1$ to finish this controlled movement.
If the ancilla
$p_1$ is in state $|e\rangle$, we erase the excitation at site $x_{2}$ by the symmetrized gauge transformation
$A(s_2)$ Eq. \eqref{eq:GT} to cancel the movement. If the ancilla $p_1$ is in state $|h\rangle$, we erase the excitation at site $x_1$ by
the symmetrized gauge transformation
$A(s_1)$ in Eq. (\ref{eq:GT}) to finish the movement.
Assume the magnetic flux at site $x_{1}$ is $\mu_{1}^{-1}$
which is the local
degree of freedom of the ancillary anyon, the controlled movement from site $x_2$ to $x_3$ is done by applying the operation
$|e\rangle_{p_{1}}\langle e|\otimes I([s_{2},s_{2}'])+|h\rangle_{p_{1}}\langle h|\otimes L_{+}^{\mu}([s_{2},s_{2}'])$
between ancilla $p_{1}$ and system qudit $[s_{2},s_{2}']$. The controlled movement back is similar. Hence, the controlled braiding can be carried out in the following steps:
\begin{enumerate}

  \item Perform the projection operation $|e\rangle\langle e|$ on the qudit on edge $[s_{1},s_{2}]$.
  \item Perform controlled unitary gate $|h\rangle_{p_1}\langle h|\otimes F(s_1)+|e\rangle_{p_1}\langle e|\otimes 1_{|G|}$
  between ancilla $p_1$ and ancilla
  $s_1$ followed by controlled unitary gate $|e\rangle_{p_1}\langle e|\otimes F(s_2)+|h\rangle_{p_1}\langle h|\otimes 1_{|G|}$
  between ancilla $p_1$ and ancilla $s_2$, where $F$ is the Fourier transformation.
  Apply the controlled gauge transformations $\sum_{g}|g\rangle_{s_{1}}\langle g|\otimes A_{g}(s_{1})$
  and $\sum_{g}|g\rangle_{s_{2}}\langle g|\otimes A_{g}(s_{2})$, measure ancillary qudits $s_{1}$ and $s_{2}$ in basis
  $|k_{[G]}\rangle$. For outcomes $k_{s_{1}}$ and $k_{s_{2}}$, perform controlled two-qudit gate
  $|e\rangle_{p_{1}}\langle e|\otimes Z_{[G]}^{k_{s_{2}}}([s_{1},s_{2}])+|h\rangle_{p_{1}}\langle h|\otimes Z_{[G]}^{k_{s_{1}}}([s_{1},s_{2}])$
  between ancilla $p_{1}$ and system qudit $[s_{1},s_{2}]$

  \item Perform controlled two-qudit gate
  $|e\rangle_{p_{1}}\langle e|\otimes I([s_{2},s_{2}'])+|h\rangle_{p_{1}}\langle h|\otimes L_{+}^{\mu}([s_{2},s_{2}'])$
  between ancilla $p_{1}$ and system qudit $[s_{2},s_{2}']$.

  \item Braid the quasi-particle at site $x_{3}$ around site $x_{i}$.

  \item Apply $|e\rangle_{p_{1}}\langle e|\otimes I([s_{2},s_{2}'])+|h\rangle_{p_{1}}\langle h|\otimes L_{+}^{\mu^{-1}}([s_{2},s_{2}'])$
  between ancilla $p_{1}$ and qudit $[s_{2},s_{2}']$, measure qudit $[s_{1},s_{2}]$ in basis $|g\rangle:g\in G$. For
  outcome $g_{1}$, perform controlled gauge transformation
  $|e\rangle_{p_{1}}\langle e|\otimes A_{g_{1}^{-1}}(s_{2})+|h\rangle_{p_{1}}\langle h|\otimes A_{g_{1}}(s_{1})$, then
  perform $A(s_{2})$.

\end{enumerate}

In order to prevent unwanted braiding caused by noise, we must locate anyons
faraway from each other, and rely on long
braiding operators for interference measurement of the topological states. In contrast, in a fusion measurement by
closed ribbon projection, when we move two anyons close enough we can choose a very small closed ribbon and perform the
corresponding projection operator once only.
Therefore, generally speaking, the topological state interference measurement is much more expensive than the fusion measurement.

\section{Proof-of-principle demonstration of non-Abelian statistics and TQC}
Since the simulation of non-Abelian anyons is very challenging, we are interested in finding out the smallest
system that is sufficient for the demonstration of non-Abelian statistics and TQC.
Here, we consider the braiding of a pure magnetic charge around
a pure electric charge, and the
fusion of two pure electric charges. And by these two processes, braiding and fusion, one could demonstrate non-Abelian  statistics and TQC.

\begin{figure}
\includegraphics[angle=0, width=0.5\textwidth]{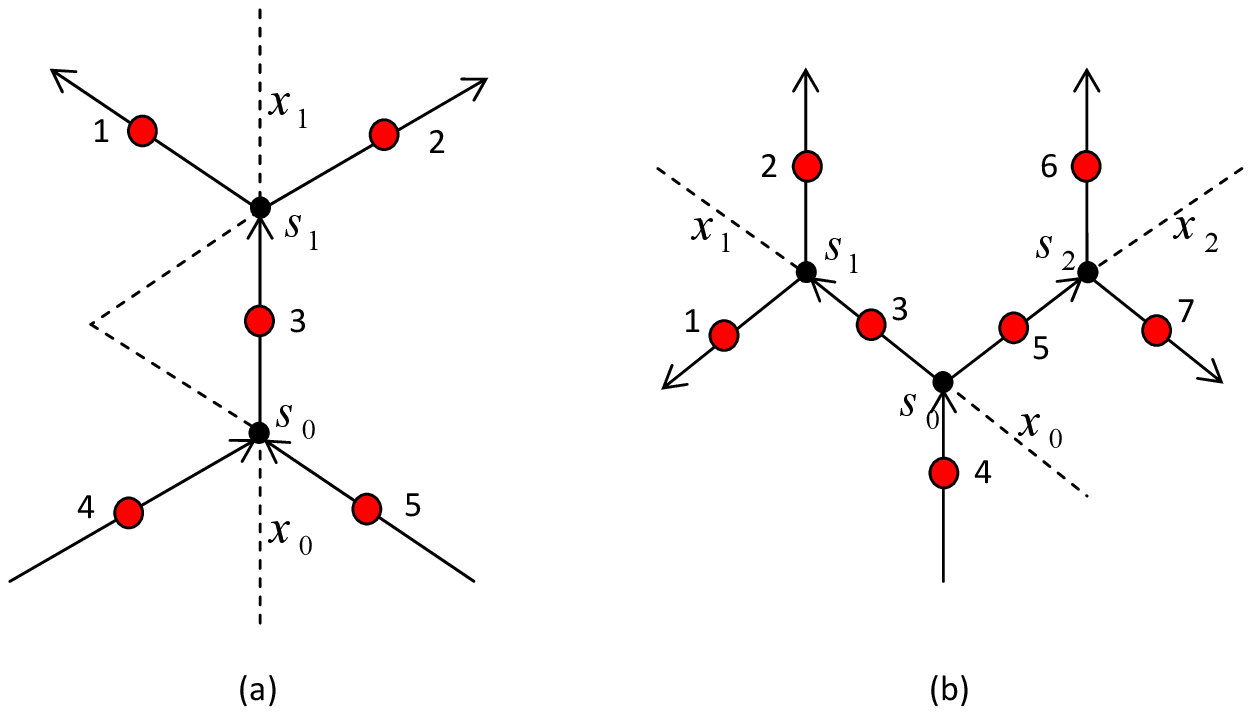}
\caption{\footnotesize{(Color online) (a)The small-scale lattice
with a minimum number of 5 system qudits (red[gray in black and
white]) for demonstration of the
 non-Abelian anyon braiding. (b) The small-scale lattice with a minimum number of 7 system qudits (red[gray in black and
white]) for the demonstration of non-Abelian anyon fusion.}}
\label{fig sscale}
\end{figure}

The graph shown in Fig. \ref{fig sscale} (a) is the smallest system for demonstration of a unitary TQC gate by anyon braiding.
We choose site
$x_{0}$ as our base site. The gauge transformation $A_{g}(s_{0})$ at $s_{0}$ is the smallest
closed ribbon operator around $s_0$. And $A_{g}(s_{0})$ can also be viewed as creating a pure magnetic charge $g$,
braiding it clockwise around
vertex $s_{0}$ and annihilating it, we denote this braiding process as $\mathcal{R}_1$.
If initially there are only pure electric charge anyons in the system,
then $\mathcal{R}_1$ corresponds to a braiding process surrounding the excitation at base site
$x_0$(pure electric charge anyons live on vertices).
Now we consider the process that one create a pure magnetic charge $g$, counterclockwise braid it around all anyonic excitations
in the system(excluding the excitation at base site $x_0$) and annihilate it, we denote this braiding process as $\mathcal{R}_2$.
So, when initially there are only pure electric charge anyons in the system, $\mathcal{R}_1\mathcal{R}_2^{-1}$
corresponds to a clockwise braiding process around all excitations in the system(including the excitation at base site $x_0$). And obviously this braiding is trivial, i.e. $\mathcal{R}_1\mathcal{R}_2^{-1}=1$ or equally
$\mathcal{R}_1=\mathcal{R}_2$.

To demonstrate the non-Abelian anyon braiding, we first prepare the system in the ground state
\begin{equation}
|GS\rangle\propto\sum_{g_1,g_2\in S_3}|g_1^{-1}\rangle_1|g_1^{-1}\rangle_2|g_1g_2^{-1}\rangle_3|g_2\rangle_4|g_2\rangle_5.
\end{equation}
Then we create a pure electric charge anyon at site $x_{1}$, or equally we can say that we create a pure electric charge
anyon at vertex $s_{1}$ (because
pure electric charge anyons live on vertices). And we braid a pure magnetic charge anyon around
this pure electric charge anyon by
applying a gauge transformation at $s_{0}$. Finally we detect the change
of the anyon state. The detection can be done by the anyonic interference method given above, we make controlled
braiding of pure magnetic charge anyons around $x_1$, but we need to repeat
the controlled braiding many times to complete the detection. Here, in this small-scale system, we can reduce the required
number of braiding to just one. Notice that the controlled braiding for detection is just the controlled gauge transformation
at $s_{0}$.
For example, we create at site $x_1$(i.e. at vertex $s_1$) the pure electric charge anyon in
topological state
\begin{equation}
\begin{split}
|0_{R_{2}}\rangle\propto&\sum_{g_1g_2^{-1}=e}|g_1^{-1}\rangle_1|g_1^{-1}\rangle_2|e\rangle_3|g_2\rangle_4|g_2\rangle_5+ \\
&\sum_{g_1g_2^{-1}=c_+}\zeta|g_1^{-1}\rangle_1|g_1^{-1}\rangle_2|c_+\rangle_3|g_2\rangle_4|g_2\rangle_5+ \\
&\sum_{g_1g_2^{-1}=c_-}\zeta^*|g_1^{-1}\rangle_1|g_1^{-1}\rangle_2|c_-\rangle_3|g_2\rangle_4|g_2\rangle_5,
\end{split}
\end{equation}
where $R_{2}$ is the two dimensional irreducible representation of
$S_{3}$(the group elements and irreps of $S_3$ are given in the appendix). Then we apply the gauge transformation
$A_{t_{0}}(s_{0})$, which realizes the braiding unitary gate
and transforms the anyon at site $x_1$(i.e. at vertex $s_1$) to topological state
$|1_{R_{2}}\rangle$ due to the braiding rules(see the appendix). To detect this change,
we prepare the ancillary qudit at $s_{0}$ in state
$\frac{1}{\sqrt{3}}(|e\rangle+|c_{+}\rangle+|c_{-}\rangle)$, apply controlled braiding
$\sum_{g}|g\rangle_{s_{0}}\langle g|\otimes A_{g}(s_{0})$, then measure the ancillary qudit $s_0$ in the Fourier basis
$\{|j\rangle=\frac{1}{\sqrt{3}}(|e\rangle+\zeta^{j}|c_{+}\rangle+\zeta^{2j}|c_{-}\rangle),j=0,1,2\}$ where $\zeta=\exp(i2\pi/3)$. If
the outcome is $|1\rangle$, the anyon state is $|0_{R_{2}}\rangle$. If the outcome is $|2\rangle$, the anyon state
is $|1_{R_{2}}\rangle$. Else if the system is in ground state, the outcome must be $|0\rangle$. Note that the
states of qudit 1 and qudit 2 are always the same in this whole demonstration. If we take the gauge
transformation $A_{g}(s_{0})$ as a single step, then the states of qudit 4 and qudit 5 are always the same too.
Because of this, we can eliminate one of the duplicates from the physical simulator for simplification. We remove
qudit 1 and 5.

\begin{figure}
\includegraphics[angle=0, width=1.0\textwidth]{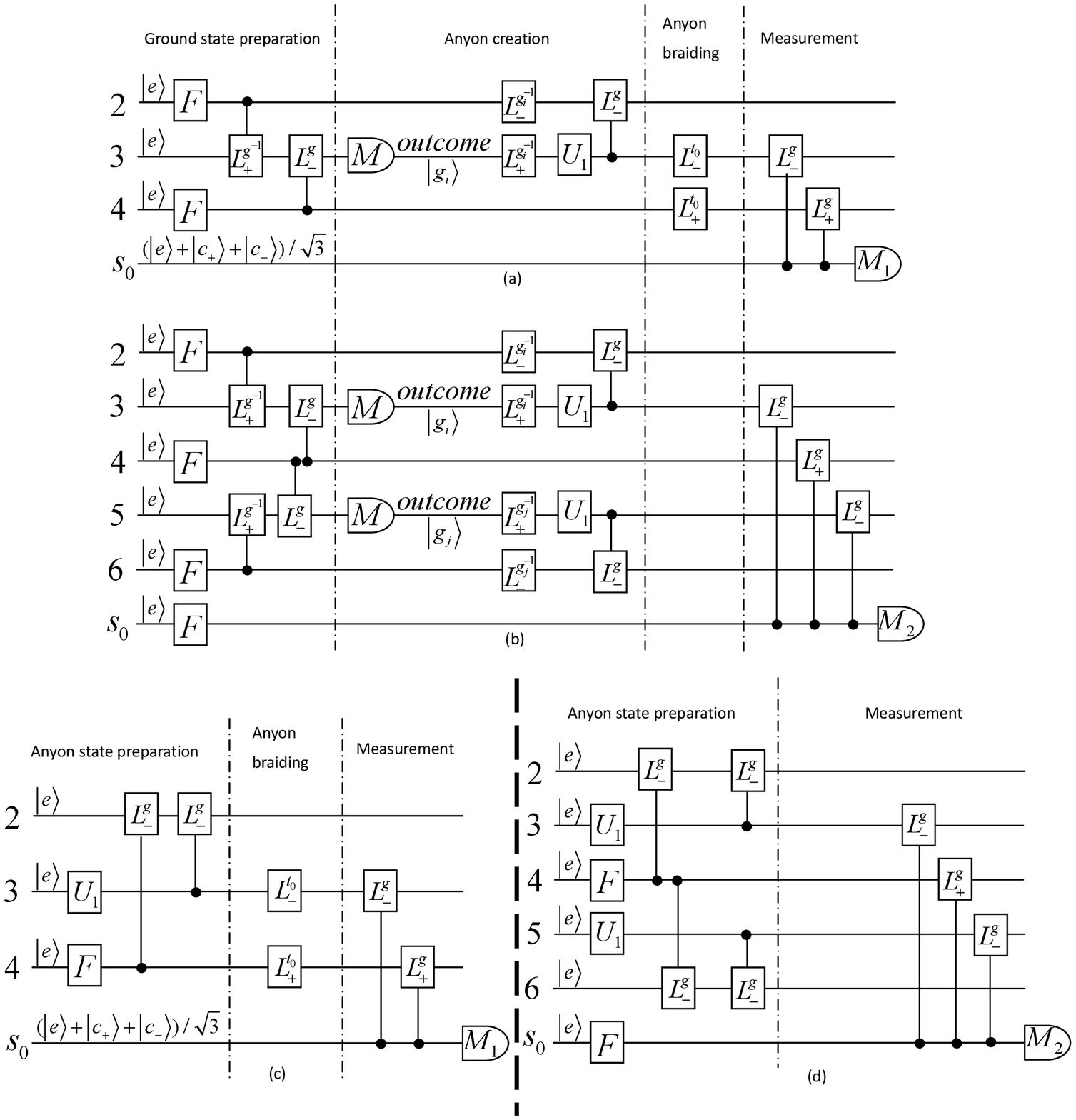}
\caption{\footnotesize{The detailed quantum circuits for the
demonstration of anyon brading and measurement. The Fourier
transformation $F$ transforms state $|e\rangle$ to
$|0_{[S_3]}\rangle$. The controlled
$U_g=\sum_{g}|g\rangle_{control}\langle g|\otimes U_{g}(target)$.
$U_1$ is an unitary operation that maps state $|e\rangle$ to
$\frac{1}{\sqrt{3}}(|e\rangle+\zeta|c_{+}\rangle+\zeta^{*}|c_{-}\rangle)$,$\zeta=\exp(i2\pi/3)$.
$M$, $M_1$ and $M_2$ are measurements in basis $\{|g\rangle\}$,
$\{|k_{[z]}\rangle=Z_{[z]}^{k}|0_{[z]}\rangle, z=e,c_+,c_-\}$ and
$\{|k_{[S_3]}\rangle=Z_{[S_3]}^{k}|0_{[S_3]}\rangle\}$. (a) is the
quantum circuit for ground state preparation, anyon creation and
braiding, and detection. (b) is the quantum circuit for ground
state preparation, anyon creation and fusion detection. (c) and
(d) are the corresponding quantum circuits with direct anyon state
preparation.}} \label{fig qcircuit}
\end{figure}

For the small scale system in Fig. \ref{fig sscale} (a), Fig. \ref{fig qcircuit} (a) shows the
detailed circuit for
the described demonstration process. We only need one ancillary qudit $s_0$ for the measurement, so the total number of qudits for braiding demonstration is 4. Initially, all system qudits 2, 3, 4 are prepared in state $|e\rangle$, and ancilla $s_0$ is in state $\frac{1}{\sqrt{3}}(|e\rangle+|c_{+}\rangle+|c_{-}\rangle)$ for measurement. For ground state preparation, we need to perform symmetrized gauge transformation $A(s)$ Eq. \ref{eq:GT} at $s_0$ and $s_1$. Take $A(s_1)$ as an example, we can first transform qudit 2 to state $|0_{[S_3]}\rangle$ by Fourier transformation, then apply controlled unitary $\sum_g|g\rangle_2\langle g|\otimes L_+^{g^{-1}}(3)$. To create the anyon at $x_1$, we first apply the projection $|e\rangle\langle e|$ on qudit 3. Then we need to perform the gauge transformations $A_e(s_1)+\zeta A_{c_+}(s_1)+\zeta^*A_{c_-}(s_1)$, this is done by first transforming qudit 3 to state $\frac{1}{\sqrt{3}}(|e\rangle+\zeta|c_{+}\rangle+\zeta^*|c_{-}\rangle)$ and then performing $\sum_g|g\rangle_3\langle g|\otimes L_-^{g}(2)$. If we do not care about the ground state and only want to know the property
of the state with anyonic excitations, we could directly prepare the system in the excited state as shown in Fig. \ref{fig qcircuit} (c)
to further simplify our scheme. We can rewrite the topological state $|0_{R_{2}}\rangle$ as
\[
\begin{split}
|0_{R_{2}}\rangle\propto&\sum_{g}|g^{-1}\rangle_2|e\rangle_3|g\rangle_4+ \\
&\sum_{g}\zeta|g^{-1}c_{+}^{-1}\rangle_2|c_+\rangle_3|g\rangle_4+ \\
&\sum_{g}\zeta^*|g^{-1}c_{-}^{-1}\rangle_2|c_-\rangle_3|g\rangle_4.
\end{split}
\]
To create this state, we first transform the system to state
\[\frac{1}{\sqrt{3}}|e\rangle_2(|e\rangle_3+\zeta|c_{+}\rangle_3+\zeta^*|c_{-}\rangle_3)|0_{[S_3]}\rangle_4\]
by single qudit unitary, where $|0_{[S_3]}\rangle=|e\rangle+|c_+\rangle+|c_-\rangle+|t_0\rangle+|t_1\rangle+|t_2\rangle$.
Then apply the controlled unitary
$\sum_g|g\rangle_4\langle g|\otimes L_-^{g}(2)$ followed by $\sum_g|g\rangle_3\langle g|\otimes L_-^{g}(2)$.
This is shown in Fig. \ref{fig qcircuit} (c).

\begin{figure}[t]
\includegraphics[angle=0, width=0.5\textwidth]{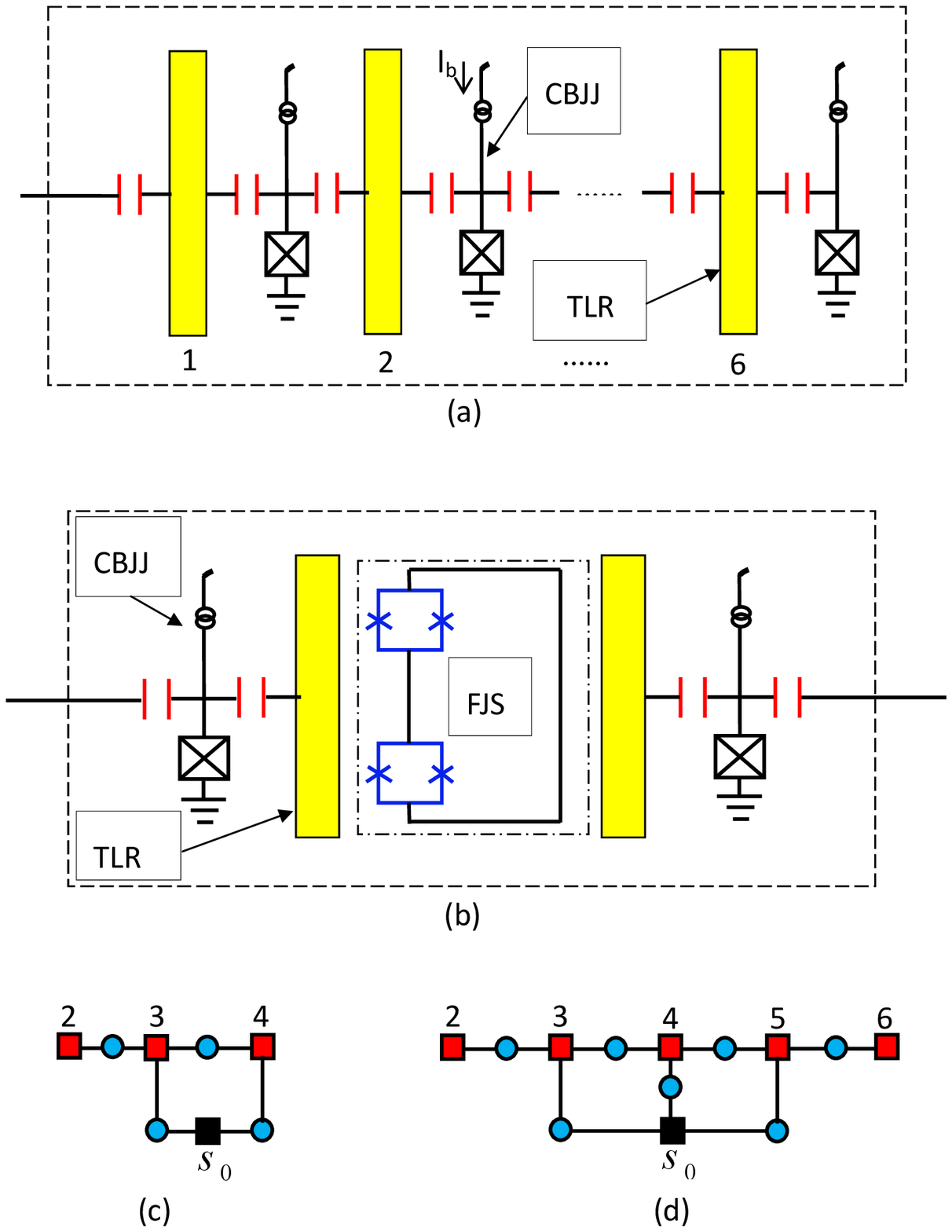}
\caption{\footnotesize{(Color online)(a) A photonic qudit based on
six TLRs that are capacitively coupled to current biased Josephson
junction.(b) A 4-junction SQUID device is used to interact
photonic qudits. (c)The superconducting circuit for demonstration
of anyons braiding statistics. (d) The superconducting circuit for
demonstration of the fusion of anyons.  System qudits (red[gray in
black and white] square) and ancillary qudits (black square) are
coupled by the 4-junction SQUID (blue[gray in black and white]
circle). }} \label{fig_circuit}
\end{figure}

For the fusion measurement demonstration, we need more sites  to place excitations.
The graph shown in Fig. \ref{fig sscale} (b) is the
smallest system for demonstration of this elementary operation, where $x_{0}$ is the base site.
We only consider the fusion measurement
of pure electric charge excitations. We first prepare the system to ground state, then create two arbitrary pure
electric charge anyons at sites
$x_{1}$ and $x_{2}$(i.e. at vertices $s_{1}$ and $s_{2}$). Since there are only the two anyons in the system, and the total topological charge of
all excitations in the system(including the excitation at base site) is vacuum.
So iff there is no excitation at the base site $x_0$,
or
equivalently iff the state of the system $|\psi\rangle$ satisfies
$A(s_0)|\psi\rangle=|\psi\rangle$(all excitations have no magnetic charge, so $B(p)\equiv1$ for all $p$),
the two pure electric charge anyons at $s_{1}$ and $s_{2}$ fuse to vacuum.
So to detect the fusion,
we prepare the ancillary qudit at $s_{0}$ in state $|0_{[S_3]}\rangle$, apply controlled gauge transformation
$\sum_{g}|g\rangle_{s_{0}}\langle g|\otimes A_{g}(s_{0})$, then measure the ancillary qudit $s_0$ in the basis
${|k_{[S_3]}\rangle=Z_{[S_3]}^{k}|0_{[S_3]}\rangle}$. If the outcome is ``$|0_{[S_3]}\rangle$'', we get vacuum after fusion. Otherwise, we get
a quasi-excitation left behind after fusion.

Since the states of qudit 1 and 2 are always the same in the whole process, so are those of
qudit 6 and 7, we can eliminate qudits 1 and 7. Also we need an ancillary qudit $s_0$ for measurement, so the total number of qudits for fusion demonstration is 6.  As an example, we can create two electric charge anyons in
topological state
$|0_{R_{2}}\rangle_{s_1}|0_{R_{2}}\rangle_{s_2}$, and measure their fusion. Based on the same consideration as the braiding demonstration, shown in Fig. \ref{fig qcircuit} (b) and Fig. \ref{fig qcircuit} (d) are the quantum
circuits for this process with and without
the ground state preparation respectively. Notice that the two anyons at $s_1$ and $s_2$ are in the same state $|0_{R_{2}}\rangle$, the operations on qudits 2,3 are the same as operations on 6,5 due to this symmetry.

Beside single qudit unitary operations on individual qudits, the above demonstration processes need two qudit gates between adjacent system qudits $i$ and $i+1$. For measurement, we need two qudit gates between ancilla $s_0$ and system qudits 3, 4, 5.

For physical implementation, we can simulate this model using a variety of physical systems such as cold atoms or trapped ions.
Here, we point out the interesting possibility of simulating it on a fully integrated
superconducting chip which provides a powerful platform for manipulating and interacting photonic qubits based on
on-chip transmission line (TLR) resonators \cite{DHZZG}. We briefly describe the system only and leave the details to elsewhere. As shown in
Fig. \ref{fig_circuit} \cite{DHZZG},
6 transmission line resonators are used as 1 qudit. There is only 1 photon in the system of these 6 resonators and its
location represents
the 6 internal states of the qudit. By coupling the TLRs capacitively to
current biased Josephson junctions (CBJJ) as shown in Fig. \ref{fig_circuit} (a), we can shift the relative energies of the TLR modes and exchange
photons between adjacent TLRs \cite{DHZZG}, and thus realize arbitrary single qudit gates. Further, we can use a four junction SQUID
shown in Fig. \ref{fig_circuit} (b) to interact photons in 2 adjacent qudits to realize the 2-qudit phase gate
$U=\exp(i\phi|g_{i}\rangle_{A}\langle g_{i}|\otimes|g_{i}\rangle_{B}\langle g_{i}|)$ \cite{DHZZG}. Thus, we can perform
all operations required to simulate the non-Abelian anyons in this system. Shown in
 Fig. \ref{fig_circuit} (c) and (d) are possible designs of the smallest superconducting circuits for the
 demonstration of anyon braiding and fusion measurement.

\section{Conclusion}
In summary, based on Kitaev's quantum double spin lattice model, we have proposed a method to simulate the non-Abelian statistics and
universal TQC by performing appropriate ribbon operators. In contrast to earlier studies, our scheme is based on braiding of anyons
created by long ribbon operators connected to a common base site, and hence can simulate genuine anyon states in a
topologically protected global space. By designing the smallest system sufficient for the demonstration of fractional statistics and
TQC, we have shown that the requirement of our proposal is only modest and it is an attractive scheme for experimental studies.

\section{Acknowledgments}
This work was funded by National Basic Research Program of China
2011CB921204, 2011CBA00200, National Natural Science Foundation of
China (Grant Nos. 60921091, 10874170, 10875110, 11174270) and the
Fundamental Research Funds for the Central Universities (Grant
Nos. WK2470000006, WK2470000004). Z. -W. Zhou gratefully
acknowledges the support of Research Fund for the Doctoral Program
of Higher Education of China (Grant No.20103402110024) and the K.
C. Wong Education Foundation, Hong Kong.

\begin{appendices}
\section*{Appendix.A.}

\chapter{\textbf{Some properties of the group$S_{3}$ :}}

$S_{3}$, the group of permutations of three objects that we label {0,1,2}, is the smallest
non-Abelian group. $S_{3}$ contains three conjugacy classes, namely:
\begin{enumerate}
\item Identity $e$.
Centralizer $N_{[e]}=S_{3}$.

\item Reflections $t_{0}=(01)$, $t_{1}=(12)$, $t_{2}=(20)$. Centralizer $N_{[t]}\cong \{e,t_{0}\}\cong \mathbb{Z}_{2}$.

\item 3-rotations $c_{+}=(012)$, $c_{-}=(021)$. Centralizer $N_{[c]}\cong\{e,c_{+},c_-\}\cong\mathbb{Z}_{3}$.

\end{enumerate}
The multiplication rules for $S_3$ are as follows:
\[
\begin{split}
t_{i}t_i &= e, t_{j}t_k = c_{\varepsilon_{j,k}}, for j\neq k.\\
t_{i}c_{\pm} &= t_{i\pm 1}, c_{\pm}t_{i} = t_{i\mp1},\\
c_{\rho}c_{\rho} &= c_{-\rho}, c_{\rho}c_{\sigma} = e, for \sigma\neq\rho.
\end{split}
\]
The operations involving $e$ are trivial. Here $\varepsilon_{j,k}=\pm$ for $k=j\pm1$(modulo 3) respectively.

The group has three irreducible representations (irreps). Two one-dimensional
irreps are the trivial one $R_{1}^{+}=1$, and the signature representation
$R_{1}^{-}(e)=R_{1}^{-}(c_{\rho})=+1$, $R_{1}^{-}(t_{i})=-1$. The two dimensional representations
are
\[
R_{2}(e)=\textbf{1}_{2}, R_{2}(t_{k})=\sigma^{x}exp(i\frac{2\pi}{3}k\sigma^z), R_{2}(c_{\pm})=exp(\pm i\frac{2\pi}{3}\sigma^z).
\]
Explicitly,
\[
R_{2}(e)=
\left(
  \begin{array}{ccc}
    1 & 0 \\
    0 & 1 \\
  \end{array}
\right),
R_{2}(c_{+})=
\left(
  \begin{array}{ccc}
    \xi & 0 \\
    0 & \xi^* \\
  \end{array}
\right),
R_{2}(c_-)=
\left(
  \begin{array}{ccc}
    \xi^* & 0 \\
    0 & \xi \\
  \end{array}
\right),
\]
\[
R_{2}(t_0)=
\left(
  \begin{array}{ccc}
    0 & 1 \\
    1 & 0 \\
  \end{array}
\right),
R_{2}(t_1)=
\left(
  \begin{array}{ccc}
    0 & \xi^* \\
    \xi & 0 \\
  \end{array}
\right),
R_{2}(t_2)=
\left(
  \begin{array}{ccc}
    0 & \xi \\
    \xi^* & 0 \\
  \end{array}
\right),
\]
where $\xi=e^{i2\pi/3}$.

The characters $\chi_{R}(g)=tr[R(g)]$ are equal to $\pm1$ for one dimensional reps. For
two dimensional reps,
\[
\chi_{R_{2}}(e)=2, \chi_{R_{2}}(t_{i})=0, \chi_{R_{2}}(c_\rho)=-1.
\]
The permutation representation for $S_3$ is a set of $6\times6$ matrices that faithfully represents group left action on the basis $\{|e\rangle, |t_0\rangle, |t_1\rangle, |t_2\rangle, |c_+\rangle, |c_-\rangle\}$, i.e. $L_+^h|g\rangle=|hg\rangle$. Similarly the right multiplication $L_-^h|g\rangle=|gh^{-1}\rangle$, $[L_+^h,L_-^{h'}]=0$ and the unitary matrices are given by
\[
L_+^e =
\left(
  \begin{array}{cccccc}
    1 & 0 & 0 & 0 & 0 & 0 \\
    0 & 1 & 0 & 0 & 0 & 0 \\
    0 & 0 & 1 & 0 & 0 & 0 \\
    0 & 0 & 0 & 1 & 0 & 0 \\
    0 & 0 & 0 & 0 & 1 & 0 \\
    0 & 0 & 0 & 0 & 0 & 1 \\
  \end{array}
\right),
L_+^{t_0} =
\left(
  \begin{array}{cccccc}
    0 & 1 & 0 & 0 & 0 & 0 \\
    1 & 0 & 0 & 0 & 0 & 0 \\
    0 & 0 & 0 & 0 & 1 & 0 \\
    0 & 0 & 0 & 0 & 0 & 1 \\
    0 & 0 & 1 & 0 & 0 & 0 \\
    0 & 0 & 0 & 1 & 0 & 0 \\
  \end{array}
\right),
\]
\[
L_+^{t_1} =
\left(
  \begin{array}{cccccc}
    0 & 0 & 1 & 0 & 0 & 0 \\
    0 & 0 & 0 & 0 & 0 & 1 \\
    1 & 0 & 0 & 0 & 0 & 0 \\
    0 & 0 & 0 & 0 & 1 & 0 \\
    0 & 0 & 0 & 1 & 0 & 0 \\
    0 & 1 & 0 & 0 & 0 & 0 \\
  \end{array}
\right),
L_+^{t_2} =
\left(
  \begin{array}{cccccc}
    0 & 0 & 0 & 1 & 0 & 0 \\
    0 & 0 & 0 & 0 & 1 & 0 \\
    0 & 0 & 0 & 0 & 0 & 1 \\
    1 & 0 & 0 & 0 & 0 & 0 \\
    0 & 1 & 0 & 0 & 0 & 0 \\
    0 & 0 & 1 & 0 & 0 & 0 \\
  \end{array}
\right),
\]
\[
L_+^{c_+} =
\left(
  \begin{array}{cccccc}
    0 & 0 & 0 & 0 & 0 & 1 \\
    0 & 0 & 1 & 0 & 0 & 0 \\
    0 & 0 & 0 & 1 & 0 & 0 \\
    0 & 1 & 0 & 0 & 0 & 0 \\
    1 & 0 & 0 & 0 & 0 & 0 \\
    0 & 0 & 0 & 0 & 1 & 0 \\
  \end{array}
\right),
L_+^{c_-} =
\left(
  \begin{array}{cccccc}
    0 & 0 & 0 & 0 & 1 & 0 \\
    0 & 0 & 0 & 1 & 0 & 0 \\
    0 & 1 & 0 & 0 & 0 & 0 \\
    0 & 0 & 1 & 0 & 0 & 0 \\
    0 & 0 & 0 & 0 & 0 & 1 \\
    1 & 0 & 0 & 0 & 0 & 0 \\
  \end{array}
\right),
\]
\[
L_-^e =
\left(
  \begin{array}{cccccc}
    1 & 0 & 0 & 0 & 0 & 0 \\
    0 & 1 & 0 & 0 & 0 & 0 \\
    0 & 0 & 1 & 0 & 0 & 0 \\
    0 & 0 & 0 & 1 & 0 & 0 \\
    0 & 0 & 0 & 0 & 1 & 0 \\
    0 & 0 & 0 & 0 & 0 & 1 \\
  \end{array}
\right),
L_-^{t_0} =
\left(
  \begin{array}{cccccc}
    0 & 1 & 0 & 0 & 0 & 0 \\
    1 & 0 & 0 & 0 & 0 & 0 \\
    0 & 0 & 0 & 0 & 0 & 1 \\
    0 & 0 & 0 & 0 & 1 & 0 \\
    0 & 0 & 0 & 1 & 0 & 0 \\
    0 & 0 & 1 & 0 & 0 & 0 \\
  \end{array}
\right),
\]
\[
L_-^{t_1} =
\left(
  \begin{array}{cccccc}
    0 & 0 & 1 & 0 & 0 & 0 \\
    0 & 0 & 0 & 0 & 1 & 0 \\
    1 & 0 & 0 & 0 & 0 & 0 \\
    0 & 0 & 0 & 0 & 0 & 1 \\
    0 & 1 & 0 & 0 & 0 & 0 \\
    0 & 0 & 0 & 1 & 0 & 0 \\
  \end{array}
\right),
L_-^{t_2} =
\left(
  \begin{array}{cccccc}
    0 & 0 & 0 & 1 & 0 & 0 \\
    0 & 0 & 0 & 0 & 0 & 1 \\
    0 & 0 & 0 & 0 & 1 & 0 \\
    1 & 0 & 0 & 0 & 0 & 0 \\
    0 & 0 & 1 & 0 & 0 & 0 \\
    0 & 1 & 0 & 0 & 0 & 0 \\
  \end{array}
\right),
\]
\[
L_-^{c_+} =
\left(
  \begin{array}{cccccc}
    0 & 0 & 0 & 0 & 1 & 0 \\
    0 & 0 & 1 & 0 & 0 & 0 \\
    0 & 0 & 0 & 1 & 0 & 0 \\
    0 & 1 & 0 & 0 & 0 & 0 \\
    0 & 0 & 0 & 0 & 0 & 1 \\
    1 & 0 & 0 & 0 & 0 & 0 \\
  \end{array}
\right),
L_-^{c_-} =
\left(
  \begin{array}{cccccc}
    0 & 0 & 0 & 0 & 0 & 1 \\
    0 & 0 & 0 & 1 & 0 & 0 \\
    0 & 1 & 0 & 0 & 0 & 0 \\
    0 & 0 & 1 & 0 & 0 & 0 \\
    1 & 0 & 0 & 0 & 0 & 0 \\
    0 & 0 & 0 & 0 & 1 & 0 \\
  \end{array}
\right).
\]
Further, the group $S_3$ has a semi-direct product structure which may be exploited to simplify physical realizations.
We have $S_3\cong\mathbb{Z}_3\rtimes_\phi\mathbb{Z}_2=\{a,b|a^3=e,b^2=e,\phi:\phi_b(a)=bab^{-1}=a^2\}$.
We can choose $\mathbb{Z}_{3}=\{e,c_{+},c_-\}$, $\mathbb{Z}_{2}=\{e,t_{0}\}$, and any element $g\in S_3$ can
be written as $g=c_{+}^{r}t_0^s$ for $r\in\{0,1,2\}, s\in\{0,1\}$. The above unitary matrices can be
simplified in $\{|r\rangle|s\rangle\equiv|c_{+}^{r}t_0^{s}\rangle\}$, a product basis for a
qutrit and qubit ~\cite{ABVC2}.

\[
\begin{split}
L_+^e=&1_3\otimes1_2, L_+^{t_0}=F(1,2)\otimes\sigma^x, L_+^{t_1}=F(0,2)\otimes\sigma^x,  \\ L_+^{t_2}=&F(0,1)\otimes\sigma^x,
L_+^{c_+}=X^{-1}\otimes1_{2},
L_+^{c_-}=X\otimes1_2, \\
L_-^{e}=&1_3\otimes1_2,
L_-^{t_0}=1_3\otimes\sigma^x, \\
L_-^{t_1}=&X^{-1}\otimes\sigma^-+X\otimes\sigma^+, L_-^{t_1}=X^{-1}\otimes\sigma^++X\otimes\sigma^-,\\
L_-^{c_+}=&X^{-1}\otimes|0\rangle\langle0|+X\otimes|1\rangle\langle 1|, \\ L_-^{c_-}=&X\otimes|0\rangle\langle0|+X^{-1}\otimes|1\rangle\langle 1|,
\end{split}
\]
where $F(i,j)=(|i\rangle\langle j|+|j\rangle\langle i|)\oplus1$ flip two basis states of qutrit,
and $X=\sum_i|i+1\rangle\langle i|$.

\section*{Appendix.B.}

\chapter{\textbf{ Some properties of the $D(S_{3}$) anyons :}}

The types of the quasi-particle of the Quantum Double model have a 1-to-1 correspondence with irreducible
representations of $D(G)$. The 8 irreps for $D(S_3)$ and the corresponding quantum dimensions are:

\begin{itemize}

 \item Vacuum $\Pi^{[e]}_{R_1^+},  d=1$;

 \item Pure electric charges $\Pi^{[e]}_{R_1^-},  \Pi^{[e]}_{R_2},  d=1, 2$;

 \item Pure magnetic charges $\Pi^{[c]}_{\beta_0},  \Pi^{[t]}_{\gamma_0},  d=2, 3$;

 \item Dyonic combinations $\Pi^{[c]}_{\beta_1},  \Pi^{[c]}_{\beta_2}, \Pi^{[t]}_{\gamma_1},  d=2, 2, 3$,\\

\end{itemize}
where $\gamma_0, \gamma_1$ correspond to the identity and signature representation of
$N_{[t]}\cong \{e,t_{0}\}\cong \mathbb{Z}_{2}$. $\beta_0,\beta_1,\beta_2$ are the three one dimensional
irreps of $N_{[c]}\cong\{e,c_{+},c_-\}\cong\mathbb{Z}_{3}$. $\beta_0$ is the identity representation.
Here we only give braiding rules of the pure charge
excitations, a full description can be found in ~\cite{A.K1,Wild}.
\[ \label{eq:braid1}
\begin{split}
\mathcal{R}&|\nu_{1}\rangle|\nu_{2}\rangle= |\nu_{1}\nu_{2}\nu_{1}^{-1}\rangle|\nu_{1}\rangle, \\
\mathcal{R}^{2}&|\nu\rangle|\xi_{n}^{R}\rangle= |\nu\rangle|R(\nu)_{mn}\xi_{m}^{R}\rangle,
\end{split}
\]
where $R$ is the unitary irrep of $S_{3}$ and $\xi_n^R$ corresponding to pure electric charge excitation,
$\nu$ corresponds to pure magnetic charge, $\mathcal{R}$ represents the counterclockwise
exchange of the two anyonic excitations. The exchange between pure electric charges is trivial.
\end{appendices}


\end{document}